GÉRALDINE MARTIN, FRANÇOISE DÉTIENNE, ELISABETH LAVIGNE

# CONFRONTATION OF VIEWPOINTS IN A CONCURRENT ENGINEERING PROCESS

**Abstract.** We present an empirical study aimed at analysing the use of viewpoints in an industrial Concurrent Engineering context. Our focus is on the viewpoints expressed in the argumentative process taking place in evaluation meetings. Our results show that arguments enabling a viewpoint or proposal to be defended are often characterized by the use of constraints. One result involved the way in which the proposals for solutions are assessed during these meetings. We have revealed the existence of specific assessment modes in these meetings as well as their combination. Then, we show that, even if some constraints are apparently identically used by the different specialists involved in meetings, various meanings and weightings are associated with these constraints by these different specialists.

1. PURPOSE

In new design and production organizations, design is often the work of a multi-speciality, multi-location team, manoeuvring, according to the moment, with the same aim (co-design) or different aims (distributed design). In the collective design process, co-design phases are specifically devoted to the assessment of the global solution, integrating the solutions produced by the different designers at time t, or to the assessment, by his/her peers, of a solution produced by one designer at time t.

A first study was focused on the coordination processes in distributed design (Martin, Détienne & Lavigne, 1999). The aim of the study presented in this paper is to analyse the viewpoints brought into play in co-design. The chosen design context is a Concurrent Engineering process. This framework seemed to us to be the most relevant for studying the topic of " viewpoint ", as the simultaneousness and confrontation of viewpoints during the development of the solution are assumed to be favoured by working in Concurrent Engineering (Darses, 1997).

Aerospatiale Matra Airbus has conducted the re-engineering of its design processes in a Concurrent Engineering procedure, in order to better master costs, schedules and quality in the design of its products. This industrial development is assisted by cognitive ergonomics research work, which is the framework of this study. We are analysing this setting up of a Concurrent Engineering methodology. The industrial aim is to derive ergonomic recommendations at software level (digital mock-up, technical database) and organizational level (meeting methodology, definition of roles) in order to assist the confrontation and integration of viewpoints in multi-speciality design.

After a brief presentation of our theoretical framework and working hypotheses, we present an empirical study aimed at analysing the use of viewpoints in an industrial Concurrent Engineering context. Our approach is strongly oriented by cognitive ergonomics work on the notion of constraint, and linguistics work on argumentation.

## 2. THEORETICAL FRAMEWORK AND WORKING HYPOTHESES

The confrontation of knowledge and the integration of viewpoints is at the heart of the cooperative mechanisms implemented in co-design. A new research topic is to characterize the viewpoints of the various players involved in collective design (designers themselves, and production and maintenance specialities) and the cooperative modes that enable these different viewpoints to be integrated.

During the design process, different viewpoints are implemented. On the basis of the work performed in different disciplines - Artificial Intelligence (Wenger, 1987), cognitive ergonomics (Rasmussen, 1979; Darses, 1997), ethnomethodology (Bucciarelli, 1998), Computer- Supported- Cooperative Work (Schmidt, 1994), an initial general definition of the notion of "viewpoint" would be : " for a person, a particular, personal, representation of an object to be designed". We are now going to develop this definition further.

In the representation of the object to be designed, and also of its design, design constraints seem to us to play a predominant part. For design problems, the solutions are not unique and correct but various, and more or less satisfactory according to the constraints that are considered. The designers develop and assess design solutions partly according to their own specific constraints, which reflect their own specific viewpoints, in relation with the specificity of the tasks they perform and their personal preferences (Eastman,1969; Falzon et al,1990).

Constraints are cognitive invariants which intervene during the design process. The notion of constraints has been understood from different angles (1) according to their origin - prescribed constraints, constructed constraints, deduced constraints, (2) according to their level of abstraction, and (3) according to their importance – validity constraints and preference constraints (Bonnardel,1999, Eastman, 1969). Futhermore, Bonnardel distinguishes various relationships between constraints.

The use of particular combination of constraints, characterizing a viewpoint, will also determine the level of abstraction at which the design object is represented. The representation of the object to be designed is characterized according to an abstract-concrete line or abstraction hierarchy (Rasmussen,1979). The different levels of abstraction are integrated into each state of the solution. This can reflect functional, structural or physical representations all along the design process (Darses,1997; Darses & Sauvagnac, 1997). Factors such as the field of expertise and specific technical interest play a role in this representation. Indeed, several participants see the design object differently according to the specificities and constraints specific to their speciality. In addition, for the same speciality, the representation will be variable according to the problem to be solved.

So, our approach is based on the following assumption: a viewpoint is (1) specific to each speciality ; (2) dependent on the problem to be solved ; 3) characterized by a level of abstraction, i.e., functional, structural or physical, (4) characterized by the implementation of a certain combination of constraints.

Our working hypothesis is that viewpoints are expressed, more or less explicitly, in multi-speciality meetings, aimed at co-design, in particular, the assessment of solutions. It is thus on the analysis of these meetings that we have focussed our empirical work.

In design activities, the assessment intervenes (1) to appreciate the suitability of partial solutions to the usual state of resolution of the problem, and (2) to select one of the solutions envisaged (Bonnardel,1999). The finality of this assessment is to make the decision to change one of its components, or to pursue the design if the assessment is positive (Darses, 1994). It is in assessment meetings that we should observe the confrontation of the viewpoints of the various participants in design. Owing to the collective nature of the activity, viewpoints should be expressed, more or less explicitly, through argumentation (Plantin, 1996). In the argumentative dialogue, a proposer will express a viewpoint that will be argued about by presenting a certain amount of information substantiating the initial proposal.

## 3. METHODOLOGY

*3.1 Context*

We conducted this study during the definition phase of an aeronautical design project, lasting three years, in which the participants work in Concurrent Engineering to design the centre section of an aircraft. These participants use Computer Assisted Design (CAD) tools and a technical Data Management System (PDM). About 400 people with 10 different specialities are involved. These specialities are the traditional design office specialities (structure, system installation, stressing), specialities that used to intervene further downstream (maintainability, production) and new specialities that have appeared with the introduction of CAD and PDM tools..

*3.2 Collection of data*

All the specialities work on the same part of the aircraft but each person according to his technical competence. "Informal" inter-speciality meetings are organized, as needed, to assess the integration of the solutions of each speciality into a global solution. We took part in 7 of these meetings as observers:
- Five meetings involved upstream design office players (designers from structure and systems installation specialits);
- Two meetings involved upstream-design office and dowstream players (from production or maintenance specialities).

On the basis of audio recordings and notes taken during the meeting, we retranscribed the full content of the meetings. Each meeting involved 3 to 6 players.

We conducted interviews afterwards with the various participants of meetings to validate the coding we had made and make explicit a certain amount of information that was implicit in the meetings.

Our second concern was to identify what representation each specialist had about constraints: in particular the representation of the meaning assigned to a constraint expressed a certain way and the ordering between constraints. For each meeting, we collected the constraints used (either explicitly or implicitly) and presented the list to each participant of this meeting. Our question concerned:
- for each constraint: to give their meaning;
- for all constraints: to order them as a function of their importance in this design-problem-situation.

*3.3 Coding scheme*

The protocols resulting from the retranscriptions were broken down according to the change of locuters. Each individual participant statements correspond to a "turn". Each turn was coded according to the following coding scheme and broken down again as required to code finer units. Our coding scheme comprises two levels, a functional level and an argumentative level.
- The functional level highlights the way in which collective design is performed. Each unit is coded by a mode (request/assertion), an action (e.g., assess) and an object (e.g., solution n). At this level, a turn can be broken down into finer units according to whether there is a change in mode, activity or object.
- The argumentative level brings out the structure of the speech on the basis of a dialogue situation. We coded the proposals for solutions made and the different types of arguments used by the speakers during the meetings.

## 4   RESULTS

Our results concerns the assessment modes, their temporal organization and the involvement of constraints in the viewpoints expressed through the argumentation process.

*4.1 Assessment modes and temporal organization*

On the basis of the coding of arguments, we have revealed the existence of analytical, comparative or analogical assessment modes in these meetings. This type of result is similar to the assessment modes analysed in individual design (Bonnardel, 1999). In addition, we have highlighted combined assessment modes, e.g. analytical/analogical.

We found that different assessment modes are used in the order shown in Figure1, whatever the meeting:
- Step1: Analytical assessment mode of the current solution;
- Step 2: if step 1 has not led to a consensus, comparative or/and analogical assessment is involved;

- Step 3: if step 2 has not led to a consensus, one (or several) argument(s) of authority is (are) used.

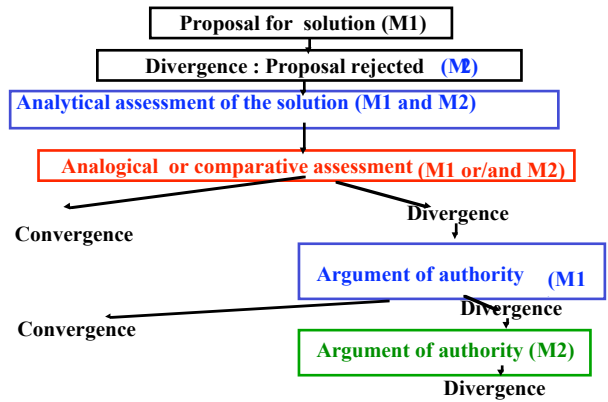

*Figure 1. The argumentation process*

Firstly the current solution is assessed. This is made using an analytical assessment mode. Arguments used by the two (or more) specialities may use more or less explicit design constraints. Specialists M1 use arguments to convince specialists M2 and M2 does the same thing. Based on this analytical assessment, a consensus is rarely found.

If no consensus has been found, then M1 and M2 use either an analogical assessment mode or a comparative assessment mode of the solution or both. The two types of assessment may also be combined. This can lead again to a consensus toward the initial solution or toward a proposed alternative solution.

Arguments used are arguments by comparison and arguments by analogy. An argument by comparison compares several objects in order to assess them. This is typically the case in the comparative assessment mode. The current solution is compared to one or several alternative solutions in function of the way they meet constraints.

An argument by analogy makes use of a precedent which is a typical case used as a model for the current case. This is typically the case in the analogical assessment mode. The current solution is compared to an analogous solution developed and already assessed in the past, either in the same project or in another project. By referring to a model of analogous reasoning the current solution is considered to be the target and the analogous solution is considered to be the source.

If no consensus has been found, either M1 or M2 propose one or several arguments of authority. Any argument can take the status of argument of authority depending on specific factors of the situation. This argument is presented as inconstestable and therefore it has a particularly strong weight in the negotiation process. We have found that an argument can take the status of argument of authority depending on :

- the status, recognised in the organisation, of the speciality that expresses it.
- the expertise of the proposer. The argument is going to make reference to a person recognized by all to be an expert in the speciality. It will be something like " It's Alphonse who said it would be more logical like that to pick up on these parts of the stringers".
- the "shared" nature of the knowledge to which it refers.

This generally leads to a consensus.

*4.2 Contraints involved in the argumentation processs*

Constraints used in the argumentation process to express viewpoints are of two kinds:
- Prescribed constraints independent of the speciality or skill-independent constraints: those constraints are prescribed in the design specification and, *a priori*, shared by all the players of the design process;
- Derived constraints specific to a speciality or skill-dependent constraints.

We found that, even though some constraints used by different players in a meeting are the same at a surface level (same terminology), these constraints may have different meanings in the viewpoints expressed by players from different specialities. Also, the level of refinement selected may be different according to the speciality.

*Selection of a meaning for a skill-independent constraint*

We observed that the same constraint (the same terms are used by different players in a meeting) can have different meanings according to the speaker's speciality.

In this case it is necessary to distinguish the two slopes of the sign, the signifier and the meaning. The meaning can have the same generic seme for different speakers but a very different functional seme. For example, a cost constraint can, for one speciality, mean "production cost" and, for another speciality, mean "design cost". It seems particularly true for general constraints prescribed for all the players of the design process (e.g., the cost) as opposed to constraints derived by a speciality (e.g., structure).

*Selection of a refinement level in a hierarchical network of a skill-dependent constraint*

We found that some constraints expressed in the argumentation process may be organized hierarchically along different levels of refinement. For example, a maintenance constraint may be refined as three constraints: accessibility constraint, dismounting constraint and mounting constraint. However, when we analysed the skill-dependent constraints used for expressing the viewpoints of different players, we identified some gaps between the level of refinement selected and used in the argumentation process according to the speaker's speciality. For a constraint specific to a skill, the level of refinement is more detailed for the speciality which represents this skill and more general for the other speciality.

*Constraints weighting*
Constraints used and their weighting, which also founds the viewpoint of the participants, depend on several factors.
- The participant's speciality;
- The interlocutors;
- The design-problem situation.

The selection of constraints depends on speaker speciality and on the interlocutors. In general, constraints taken into account in a particular meeting are those constraints specific to the specialities involved in the meeting in addition to the prescribed constraints. However skill-dependent-constraint weighting depends on speaker speciality. Whereas we found a high intra-speciality agreement on constraint weighting, we found disagreement between specialities.

For exemple, in a meeting involving Hydraulic system intallation specialists and Structure specialists, we observed that the constraints which are specific to Hydraulic system intallation specialists are : system installation and frontier. The constraints which are specific to Structure specialists are : structure and stress. Even if most of these constraints are used by the two specialities involved in the meeting, the way each speciality orders those constraints by importance is different. Each specialist ranks his/her own constraints as more important than the constraints of his/her interlocutors.

Constraints weighting also depends on the problem in hand. For example, we observed for the same speciality, air system installation, that constraint weighting varied between two problems processed sequentially in a meeting : the maintainability constraint was ranked as being of average importance for problem A and as being of high importance for problem B. Furthermore the production constraint was evoked only for problem A.

## 5   DISCUSSION AND IMPLICATIONS

This paper presents an initial empirical study of viewpoints expressed through the argumentation process in design. Our results have two kinds of implication.

We have shown that the argumentation process involves knowledge on the current solution, i.e., the solution to be assessed, but also on other solutions, i.e., alternative solutions or source solutions. This is involved in comparative assessment modes and analogical assessment modes. This result highlights the importance of documenting the design rationale for the current solutions but also those for the other solutions evoked.

We have also shown that viewpoints involve constraints which may be skill-independent or skill-dependent. The meaning and weighting of these constraints greatly depends on the multi-speciality context. This result should also be taken into account so as to better support and document the decision process in design.

## 6   REFERENCES

Bonnardel, N. (1999) L'évaluation réflexive dans la dynamique de l'activité du concepteur. In J. Perrin (Ed) : *pilotage et évaluation des processus de conception*. L'Harmattan.

Bucciarelli, L.L (1988) An ethnographic perspective on engineering design. *Design studies,  9* (3), 159-168.

Darses, F. (1994) *Gestion des contraintes dans la résolution de problèmes de Conception*. PhD Dissertation, University Paris 8, France.

Darses, F. (1997) L'ingénierie concourante : Un modèle en meilleure adéquation avec les processus cognitifs de Conception.  In P. Bossard, C.Chanchevrier & P.Leclair (Eds) *: Ingénierie concourante de la technique au social.* Economica. Paris.

Darses, F., & Sauvagnac, C. (1997) Représentations cognitives de l'objet à concevoir : construction collective dans une situation de Conception continue. *Proceedings of 01 Design 97*, Théoule-sur-mer, 24-26 septembre. Europia: Paris.

Eastman, C. M. (1969) Cognitive processes and ill-defined problems: a case study from design. In D.E. Walker & L. M. Norton (Eds*): Proceedings of the First Joint International Conference on Artificial Intelligence*. Bedford, MA: MITRE

Falzon, P., Bisseret, A., Bonnardel, N., Darses, F., Détienne, F., & Visser, W. (1990) Les activités de conception: l'approche de l'ergonomie cognitive.  *Actes du Colloque Recherches sur le design. Incitations, implications, interactions*, Compiègne, 17-19 octobre 1990.

Martin, G. , Détienne F. & Lavigne E. (1999) Le processus de conception en ingénierie concourante : une étude ergonomique. *Proceedings of MICAD 99*, Edition Hermes Science, p 215-222.

Plantin, C. (1996) *L'argumentation*. Seuil.

Rasmussen, J. (1979) *On the structure of knowledge" a morphology of mental models in a Man-Machine System context*.  Technical report Riso-m-2192, Riso national laboratory, DK-4000  Roskilde, Denmark.

Schmidt, K. (1994) *Modes and Mechanisms of interaction in Cooperative Work : outline of a conceptual framework* . Technical report Riso National Laboratory, DK-4000 Roskilde, Denmark.

Wenger, E. ( 1987) *Artificial intelligence and tutoring systems: computational and cognitive approaches to the communication of knowledge*. Morgan Kaufman Publishers. Californy, US.

## 7   AFFILIATIONS

*Françoise Détienne and Géraldine Martin,*
*EIFFEL Research Group "cognition and cooperation in design", INRIA*
*Domaine de Voluceau, Rocquencourt, BP 105, 78153 Le Chesnay, France.*
*Géraldine Martin and Elysabeth Lavigne*
*EADS AIRBUS-SA BTE/SM/GDT-CAO M0101/9 316 route de bayonne31060*
*Toulouse cedex 03, France.*